\documentclass[twocolumn,showpacs,aps,prd,amsmath,amssymb,superscriptaddress,nofootinbib,nobibnotes,floatfix]{revtex4-1}

\usepackage{graphicx}
\usepackage{float}
\usepackage{bm}
\usepackage{hyperref}
\usepackage{color}
\pdfoutput=1

\usepackage[utf8]{inputenc}
\usepackage[T1]{fontenc}
\usepackage{slashed}

\usepackage{enumerate}

\usepackage{amssymb,amsthm}
\usepackage{amsmath}

\begin{document}

\title{
   Binary neutron star mergers in Einstein-scalar-Gauss-Bonnet gravity
   }
\author{William E. East}
\email{weast@perimeterinstitute.ca}
\affiliation{%
Perimeter Institute for Theoretical Physics, Waterloo, Ontario N2L 2Y5, Canada
}%
\author{Frans Pretorius}
\email{fpretori@princeton.edu}
\affiliation{Department of Physics, Princeton University, Princeton, New Jersey 08544, USA}

\begin{abstract}
Binary neutron star mergers, which can lead to less massive black holes
relative to other known astrophysical channels, have the potential to probe
modifications to general relativity that arise at smaller curvature scales
compared to more massive compact object binaries. As a representative
example of this, here we study binary neutron star mergers in
shift-symmetric Einstein-scalar-Gauss-Bonnet gravity using evolutions of
the full, nonperturbative evolution equations.  We find that
the impact on the inspiral is small, even at large values of the modified gravity coupling (as expected,
as neutron stars do not
have scalar charge in this theory). However, postmerger
there can be strong scalar effects, including radiation. When a black hole forms, it 
develops scalar charge, impacting the ringdown gravitational wave signal.
In cases where a longer-lived remnant star persists 
postmerger, we find that the oscillations of the star source levels of scalar
radiation similar to the black hole formation cases. In remnant stars, we further find that at 
coupling values comparable to 
the maximum value for which black hole solutions of the same mass exist, there is significant nonlinear 
enhancement in the scalar field, which if sufficiently large leads to a breakdown in the evolution, seemingly
due to loss of hyperbolicity of the underlying equations.
\end{abstract}
\maketitle
\section{Introduction}%
Recent breakthroughs in gravitational wave astronomy have allowed for
unprecedented tests of general relativity (GR) in the strong field
regime~\cite{LIGOScientific:2020tif,LIGOScientific:2021sio}.  However, a
crucial step in being able to perform the most sensitive searches for
modifications to GR, or in the absence of deviations, place the most stringent
constraints, is obtaining predictions in alternative theories, in particular
in the strong field regime.

A common feature of many proposed modifications to GR is that they show the
strongest effects in the presence of the shortest curvature lengths. This is a
natural consequence of adding additional curvature terms to the Einstein-Hilbert
action multiplied by constants whose dimension are some positive powers of
length, as in dynamical Chern-Simons gravity~\cite{Alexander:2009tp}, the most generic of the
Horndeski class of theories~\cite{1974IJTP...10..363H}, or theories that add terms
constructed out of higher powers of the Riemann tensor without introducing
additional light degrees of freedom~\cite{Endlich:2017tqa,deRham:2020ejn}.  An
ideal way to look for evidence of, or to constrain, such theories is by observing
the smallest mass compact objects.

The vast majority of observed galactic black holes have masses $>5\
M_{\odot}$~\cite{Miller:2014aaa}, with the candidate lowest mass black hole
having a mass $3.3_{-0.7}^{+2.8}\ M_{\odot}$~\cite{Thompson:2018ycv}, leading
to a hypothesized so-called lower-mass gap between the highest mass neutron
star and the lowest mass black hole.  The gravitational wave event GW190814
from a binary with a $2.6\ M_{\odot}$ compact
object~\cite{LIGOScientific:2020zkf}, which could potentially be a neutron star
or black hole, has renewed debate about the lower-mass gap, though population
models currently have difficulty explaining such a low-mass black
hole~\cite{Zevin:2020gma}.  Although there are a number of speculative or exotic
formation channels that could lead to low mass black holes, one likely way to
form a black hole of mass $\sim 3\ M_{\odot}$ is from the merger of a binary
neutron star. In this work, we study how binary neutron star mergers can
be used to probe a representative modified gravity theory,
Einstein-Scalar-Gauss-Bonnet (ESGB) theory, which introduces modifications to
GR at small curvature length scales (corresponding to sufficiently high curvature).

There have been numerous studies of neutron star mergers in theories that do
not modify the principal part of the Einstein equations, in particular
scalar-tensor theories. Here, it is the introduction of a new scalar degree
of freedom that mediates a prescribed conformal rescaling of
the metric, 
rather than a modification of the Einstein equations themselves,
that can lead to novel physics. For example, neutron stars typically develop scalar charge,
which can lead to dipole radiation in a binary system containing a charged neutron
star. The lack of any observed signatures of this in binary pulsar systems
give tight constraints on such scalar-tensor theories~\cite{Damour:1996ke,Will:2014kxa}.
However, there are some notable examples where such pulsar systems
may not be strongly affected by scalar modifications at their current separations, yet where 
there could be significant modifications to the late inspiral or merger phase. 
For example, scalar-tensor theories with screening mechanisms\cite{terHaar:2020xxb,Bezares:2021yek,Bezares:2021dma}, or in
the class of scalar-tensor theories developed by Damour and Esposito-Far\`{e}se~\cite{Damour:1992we,Damour:1993hw},
where in some cases only neutrons above a certain mass can develop scalar charge
(so-called ``spontaneous scalarization''), or even only develop this
charge in the late stages of inspiral (``dynamical scalarization'')~\cite{Barausse:2012da,Palenzuela:2013hsa}.
Though the observation of a $\sim2\ M_{\odot}$ neutron star in orbit with a white dwarf severely
constrains even this class of scalar-tensor theory~\cite{Antoniadis:2013pzd}, there is
still some theoretical maneuvering that can evade these constraints, for example by
giving the scalar field a small mass~\cite{Ramazanoglu:2016kul}.

In contrast, full compact object mergers in modified theories that do change the principal part of the
Einstein equations have been less well studied, in part because of 
difficulties with finding well-posed formulations of the evolution 
equations of such theories.
In this work, we take advantage of recent advances in solving the full equations of shift-symmetric
ESGB gravity to study binary
neutron star mergers, as well as the collapse of isolated, hypermassive neutron stars to black holes.
In particular,
we use the modified harmonic formulation~\cite{Kovacs:2020pns,Kovacs:2020ywu}
and the methods developed in Ref.~\cite{East:2020hgw} for evolving
binary black holes in Horndeski theories. For a recent, detailed 
review, see Ref.~\cite{Ripley:2022cdh}.

To our knowledge, the only prior numerical study of the dynamics of neutron stars within
ESGB gravity is the work of Ref.~\cite{R:2022cwe}, where the collapse of a neutron
star to a black hole in the decoupling limit of ESGB gravity was considered (see
related earlier work in Ref.~\cite{Benkel:2016rlz} where Oppenheimer-Snyder collapse
of a pressureless fluid was examined). In the decoupling limit, the 
backreaction of the ESGB scalar is ignored and the ESGB
scalar is evolved on the pure-GR background of a collapsing neutron star spacetime.
Though this approach, as detailed in Ref.~\cite{R:2022cwe},
gives important information regarding the growth of scalar hair about
the nascent black hole, it is unable to address
at least two important questions: what the potential gravitational wave signatures
of ESGB gravity are (the scalar radiation is by itself not measurable 
with present detectors), and what the realm of validity of the small coupling 
approximation is (including what happens when this approximation is no longer valid).

Regarding potential observational signatures, an interesting aspect of ESGB
gravity is that neutron stars carry no scalar charge, yet black holes do.
(Though note, as discussed below, stationary black hole solutions only exist above a
minimum mass set by the coupling scale of the theory.) Similar to the class of
Damour-Esposito-Far\`{e}se scalar-tensor theories mentioned above, this then
implies ESGB gravity can easily evade binary pulsar system constraints, and
instead one would need to look to compact object merger dynamics to uncover
signs of it (or hope for the discovery of a galactic black hole-pulsar binary).

There has been much work constraining ESGB gravity with binaries containing one or two black holes
(see, e.g., Ref.~\cite{Nair:2019iur,Perkins:2021mhb,Lyu:2022gdr} and references therein),
with the upshot, as discussed further in Sec.~\ref{sec_id}, that they constrain
the relevant coupling length scale ($\sqrt{\alpha_{\rm GB}}$, defined below) to be on the order of a kilometer
or less.
The effect of ESGB modifications on the neutron star maximum mass and tidal deformability has also been considered~\cite{Saffer:2021gak},
though this is more difficult to separate from the unknown neutron star equation of state.
Since the smallest compact objects offer the best probes of ESGB gravity,
barring the confirmed existence of subsolar mass black holes of primordial or other exotic origin, it seems likely
that observing gravitational waves from compact object mergers will continue
to be able to place the tightest constraints on ESGB gravity. 

As the majority of theoretical work has focused on black hole binaries in
ESGB gravity~\cite{Yagi:2011xp,Witek:2018dmd,Silva:2020omi,Okounkova:2020rqw,
East:2020hgw,East:2021bqk,Shiralilou:2020gah,Shiralilou:2021mfl},
there still is an open question regarding whether binary
neutron star mergers could give comparable or better constraints
than the typical merger involving a black hole. This could either
be due to the formation of a small, scalar-charged black hole post merger,
or in the late stages of inspiral, merger, and evolution of a hypermassive neutron
star remnant, where nonlinear or strong coupling effects could be significant
(and note that, unlike with spontaneous scalarization, a neutron star in ESGB 
gravity will have a scalar cloud around it sourced by the Gauss Bonnet (GB) curvature---it
is just that this cloud falls off much more rapidly than the $1/r$ decay that would be required 
for the neutron star to register a scalar charge). 

The main goal of this paper is to begin to address the questions just posed.
Qualitatively, the answers suggested by our results are mixed in this regard.
On the optimistic side, the apparent breakdown of hyperbolicity
in the evolution for large values of the ESGB coupling suggest 
that a typical binary neutron star merger, even without assuming black hole formation, pushes shift-symmetric ESGB past the 
breaking point of theory unless $\sqrt{\alpha_{\rm GB}}$ of $\lesssim 1$ km,
comparable to the best existing constraints from mergers containing a black hole. 
Less optimistic are if one hopes to do better than this by measuring
details of the gravitational wave emission.
We find that the effects of ESGB on the gravitational wave emission show up primarily
in the postmerger signal: for a hypermassive remnant,
the oscillating high density core can excite the scalar field,
and for prompt collapse to a black hole 
the ringdown signal is affected by the development of scalar charge.
However, 
even for strong couplings close to the maximum allowed, 
these appear
sufficiently minor that it may be difficult to disentangle the effects
of departures from GR from parameter uncertainties or limited knowledge of the neutron
star equation of state (though a more quantitative analysis, beyond the scope
of this paper, would be needed for more conclusive answers).
Adding to the challenge, these parts of the gravitational wave signal are 
at high frequencies that ground-based detectors are less sensitive to.

In earlier, full nonlinear studies of collapse and black holes in ESGB
gravity~\cite{Ripley:2019hxt,Ripley:2019irj,Ripley:2020vpk,East:2020hgw,East:2021bqk,Franchini:2022ukz},
it was found that when the coupling is made too large, the hyperbolicity of the
evolution equations breaks down prior to any singular behavior developing in
the metric or scalar field.
Here we find evidence this can happen in neutron star mergers not only when a black hole
forms, but also during the postmerger oscillations of a remnant star, with apparent
breakdown in the latter occurring at comparable but somewhat larger values
of the coupling constant compared to when it does during black hole formation.
(Though unlike the spherically
symmetric studies in~\cite{Ripley:2019hxt,Ripley:2019irj}, here we do
not explicitly compute the characteristics of the full system, and only
surmise that this is the cause of the breakdown of our numerical evolutions.)
In other words, even though exceeding the weak coupling limit in ESGB gravity
has dire consequences for well-posedness of the theory,
approaching this limit in a dynamical setting does not appear to be preceded by 
novel or dramatically different spacetime/scalar field dynamics compared to 
far-from maximum coupling.

An outline of the remainder of this paper is as follows. We review shift-symmetric ESGB, the gravity
theory we consider here, in Sec.~\ref{sec:esgb}; we describe our methods for
numerically evolving this theory coupled to hydrodynamics and analyzing the
results in Sec.~\ref{sec:methods}; we present results from our study
of neutron star mergers and collapse of unstable hypermassive neutron 
stars in Sec.~\ref{sec:results}; and we discuss these
results and conclude in Sec.~\ref{sec:discuss}.
Unless otherwise noted, we use geometric units with $G=c=1$.

\section{Shift-Symmetric Einstein Scalar Gauss Bonnet}
\label{sec:esgb}
The action for shift-symmetric ESGB gravity is given by 
\begin{align}
\label{eq:action}
    S
    =
    \frac{1}{8\pi}
  \int d^4x\sqrt{-g}
\left(
        \frac{1}{2}R
    -\frac{1}{2}\left(\nabla\phi\right)^2 	
    +	\lambda\phi\mathcal{G}
    \right) + S_{\rm matter}
    ,
\end{align}
where $g$ is the determinant of the spacetime metric,
$\mathcal{G}$ is the GB scalar,
given in terms of the Riemann tensor and its contractions as
\begin{align}
    \mathcal{G}
    := 
R^2-4R^{ab}R_{ab}+R^{abcd}R_{abcd},
\end{align}
$\lambda$ is a coupling constant with dimensions of length squared,
$\phi$ is the scalar field,
and $S_{\rm matter}$ is the action for any other matter (in our case, the neutron star fluid). 
The equations of motion are given by
\begin{align}
\label{eq:eom_scalar}
      \Box\phi
   +  \lambda \mathcal{G}
    &=&
   0
   ,\\
\label{eq:eom_metric}
   R_{ab}
-  \frac{1}{2}g_{ab}R
+  2\lambda\delta^{efcd}_{ijg(a}g_{b)d}R^{ij}{}_{ef}
   \nabla^g\nabla_c \phi 
    &=&
   8\pi T_{ab} 
   ,
\end{align}
where $\delta^{abcd}_{efgh}$ is the generalized Kronecker delta and $T_{ab}=T^{\rm matter}_{ab}+T_{ab}^{\rm SF}$ with 
\begin{align}
    T_{ab}^{\rm SF}
   :=
   \frac{1}{8\pi}\left(
      \nabla_a\phi\nabla_b\phi-\frac{1}{2}g_{ab}\nabla_c \phi \nabla^c \phi
   \right) 
    \  .
\end{align}
The other matter equations of motion are not affected by the GB term, and are the same as in GR.

In this theory, stationary black holes have nonzero scalar charge $Q_{\rm SF}$.
That is, at large radius, the scalar field falls of like
$\phi=Q_{\rm SF}/r+\mathcal{O}(1/r^2)$.
Furthermore, studies have found that for a given black hole mass and spin 
there is a maximum value of $\lambda$, above which stationary solutions  
no longer exist. For a nonspinning black hole, $\lambda \lessapprox  0.23 M^2$~\cite{Sotiriou:2014pfa}, where $M$ is the total
mass, as measured at infinity, while for dimensionless black hole spins $a=0.7$ and 0.8,
$\lambda/M^2 \lessapprox 0.19$ and $0.16$, respectively~\cite{Delgado:2020rev}. 

Neutron stars, in contrast to black holes, do not have scalar charge in ESGB gravity.
Recalling the argument given in Ref.~\cite{Yagi:2011xp}, if one assumes a
stationary, asymptotically flat star solution and integrates Eq.~\eqref{eq:eom_scalar} over
the four-dimensional spacetime manifold, this gives
\begin{equation}
\int \Box\phi \sqrt{-g} d^4x = -\lambda \int \mathcal{G} \sqrt{-g} d^4x = 0 ,
\end{equation}
with the last equality following from the fact that the integral of the Gauss Bonnet (GB) curvature is a topological invariant. 
Using stationarity to drop the time integration, and applying Stoke's theorem to the
remaining spatial volume integral, we obtain a surface integral at spatial infinity contracted
with the unit normal to the surface
\begin{equation}
    \int \hat{n}^i (\partial_i \phi) \sqrt{-g} dS \propto Q_{\rm SF}=0 \ . 
\end{equation}
Note that this argument does not apply to the black hole case due to the breakdown
of the regularity of the solution in the black hole interior.

\section{Methodology\label{sec:methodology}}%
\label{sec:methods}

\subsection{Evolution}
We evolve the full, nonperturbative, shift-symmetric ESGB equations in the
modified generalized harmonic formulation~\cite{Kovacs:2020pns,Kovacs:2020ywu}
using the implementation and methods of Ref.~\cite{East:2020hgw}.
In this formulation, there are two additional auxiliary metrics $\hat{g}_{ab}$ 
and $\tilde{g}_{ab}$, which, respectively, determine
the light cone for the gauge and constraint propagating modes. As in Ref.~\cite{East:2020hgw}, we
choose $\tilde{g}_{ab}=g_{ab}-(1/5)n_an_b$ and $\hat{g}_{ab}=g_{ab}-(2/5)n_an_b$,
where $g_{ab}$ is the physical metric, and $n_a$ is the future-directed unit normal to slices
of constant time. 
The gauge we use is the modified (by the auxiliary metric) version of the damped
harmonic gauge~\cite{Lindblom:2009tu,Choptuik:2009ww}.

We model the neutron stars using ideal hydrodynamics. The Euler equations are
unmodified from the GR case (only the metric going into the equations will be
different than in GR), and we use the hydrodynamics code of
Ref.~\cite{East:2011aa} to evolve the fluid, and in particular, we use the same
methods and parameters for evolving binary neutron stars as in
Ref.~\cite{East:2019lbk}.  In the Appendix, we provide details on
the numerical resolution and convergence.

\subsection{Initial data and cases considered}\label{sec_id}
We use quasicircular binary neutron star initial data constructed with the Compact Object
CALculator~({\tt COCAL})~\cite{Tsokaros:2015fea,Tsokaros:2018dqs}.  For the
scalar field, we choose $\phi=\partial_t\phi=0$ on the initial time slice, in
which case the constraint equations of ESGB are the same as in GR.  This means
that at the beginning of the evolution there will be a short transient associated
with the scalar field evolving to a nonzero value in the presence of the neutron stars.
For the binary neutron stars, we use a piecewise polytropic form of the DD2 EOS~\cite{Banik:2014qja}.

We focus on equal mass binary neutron stars with an initial separation of 45 km,
approximately four orbital periods before merger. We consider two values for
the total mass of the system: $M=3.0\ M_{\odot}$, which gives rise to a
longer-lived hypermassive remnant; and $M=3.45\ M_{\odot}$, which promptly
collapses to a black hole postmerger.
We consider ESGB coupling parameters approaching, and in some cases exceeding,
the maximum values where our evolutions break down, which depends on whether
black hole formation occurs.
For the longer-lived remnant cases, we consider ESGB coupling parameters
$\lambda/M^2=0$, 0.04, 0.08, 0.2, 0.25, and 0.3, while for
the prompt black hole cases we consider smaller values of $\lambda/M^2=0$, 0.02, and 0.03.

We also consider the axisymmetric collapse of uniformly rotating hypermassive neutron stars.
For initial data, we use a stationary (in GR) but unstable star solution constructed using the RNS code~\cite{Stergioulas:1994ea}
with the piecewise polytropic representation of the ENG EOS~\cite{Engvik:1995gn} from~\cite{Read:2008iy}
with a mass $M=2.64\ M_{\odot}$ and a dimensionless spin of $0.7$.
The collapse of this model in GR was previously considered in Ref.~\cite{Zhang:2020qlh}.
For this scenario, we consider ESGB coupling parameters
$\lambda/M^2=0$, 0.05, 0.065, and 0.08.

For ease of comparison with other works, we convert our coupling $\lambda$ into the $\alpha_{\rm GB}:=\lambda/\sqrt{8\pi}$
used in, e.g., Refs.~\cite{Perkins:2021mhb,Lyu:2022gdr},
\footnote{Some other references~\cite{Witek:2018dmd,Blazquez-Salcedo:2016enn,Pierini:2021jxd,Pierini:2022eim} use a convention that gives a value of $\alpha_{\rm GB}$
that is $16\sqrt{\pi}$ times higher.}
 and restore physical units. 
We have that 
\begin{equation}
    \sqrt{\alpha_{\rm GB}}\approx 1.98 \ {\rm km} \left(\frac{\lambda^{1/2}}{M}\right)\left(\frac{M}{3 \ M_{\odot}}\right) \ .
\end{equation}
For reference, in Ref.~\cite{Lyu:2022gdr}, a constraint of $\sqrt{\alpha_{\rm GB}} \lesssim
1.2$ km (90\% confidence level) is found by comparing several black hole-neutron star
and binary black hole gravitational wave signals to post-Newtonian results for ESGB.

\subsection{Diagnostic quantities}
To determine the gravitational wave signal, we compute the Newman-Penrose scalar $\Psi_4$
on coordinate spheres at large radii ($r=100M$), and decompose this quantity into
spin $-2$ weighted spherical harmonics. 

In addition to the gravitational waves, we also analyze several
quantities related to the scalar field. 
Considering just the
canonical scalar stress-energy tensor,
we calculate several associated quantities, including 
the associated energy 
\begin{align}
    E_{\rm SF} 
   :=
    -\int \left(T_{t}^t\right)^{\rm SF} \sqrt{-g}d^3x \ ,
\end{align}
and energy flux through a surface in the wavezone 
\begin{align}
    \dot{E}_{\rm SF}
   \equiv
    -\int \alpha \left(T_t^i\right)^{\rm SF} dA_i \ ,
\end{align}
where $\alpha$ is the lapse.
We note that $T_{ab}^{\rm SF}$ is not conserved, and, for example, even for an 
isolated black hole with scalar charge in ESGB, $E_{\rm SF}$ will only account
for a fraction of the difference between the global mass and black hole horizon
mass. 
We also consider the value of $\phi$ on
a sphere at large radius $r=100M$, using its average value to calculate the scalar charge, 
as well as calculating the value of other (spin 0) spherical harmonics.

\section{Results}
\label{sec:results}
We follow the evolution of three different scenarios: a binary neutron star
that promptly collapses to a black hole after merger, a binary neutron star
that forms a massive remnant star at merger, and the collapse of an unstable
uniformly rotating hypermassive neutron star.  The last mentioned case
approximates the scenario where a postmerger remnant star collapses to a black
hole on long time scales (on the order of 100 ms~\cite{Hotokezaka:2013iia}), after sufficient cooling and the dissipation of
differential rotation.  For all these scenarios, we vary the ESGB coupling
$\alpha_{\rm GB}$ all the way up to near the maximum value where we are able to
carry out the evolution, and analyze the impact on the gravitational wave and
scalar radiation.

\begin{figure}
    \centering
    \includegraphics[width=\columnwidth,draft=false]{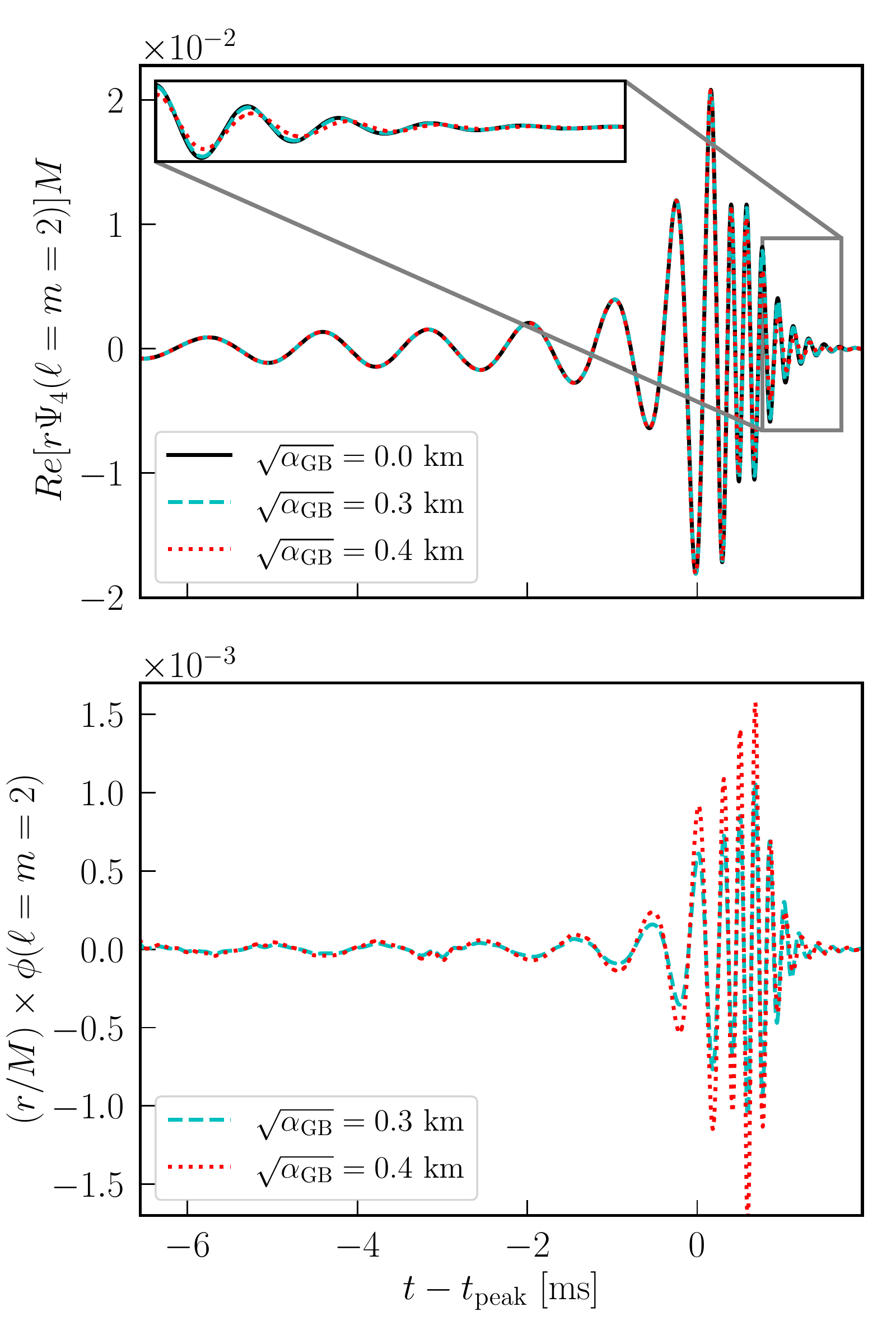}
\caption{
    Gravitational wave radiation (top) and scalar radiation (bottom) for a
    binary neutron star merger that promptly collapses to black hole.  In
    particular, we show the real part of the $\ell=m=2$ spherical harmonic of the Newman-Penrose
    scalar $\Psi_4$ and $\phi$.  The inset in the top panel shows the small
    differences during the ringdown.  Time is measured in milliseconds with
    respect to the time when the gravitational wave luminosity is maximum $t_{\rm peak}$.
\label{fig:collapse_rad}
}
\end{figure}

\begin{figure}
    \centering
    \includegraphics[width=\columnwidth,draft=false]{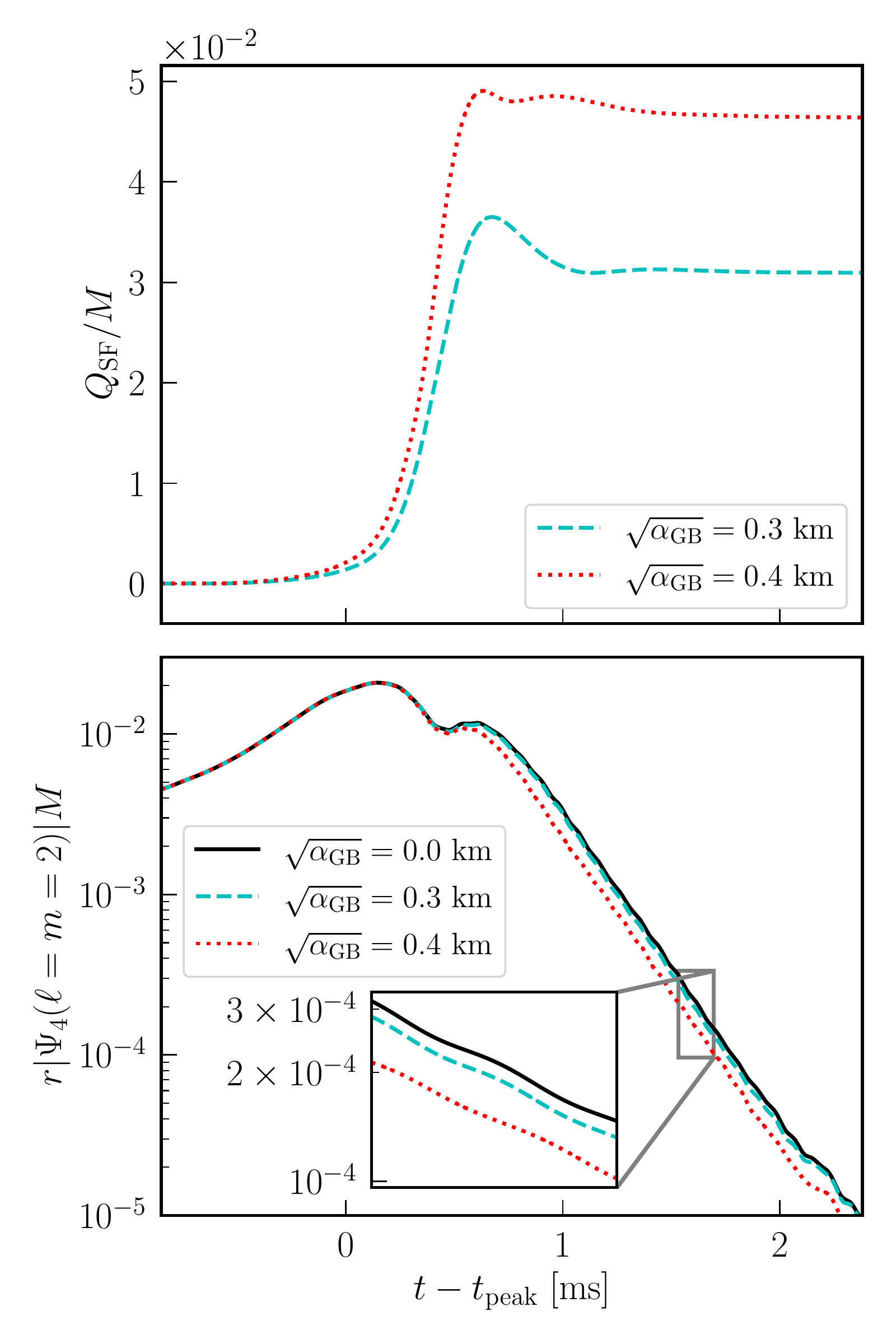}
\caption{
    Top: The scalar charge $Q_{\rm SF}$, as measured from the average scalar field
    value at large distances, as a function of time, for the binary neutron
    star mergers that promptly collapse to a black hole.  
    Bottom: The amplitude of the $\ell=m=2$ spherical harmonic $\Psi_4$ for the same time interval.
    During the black hole ringdown, the $\sqrt{\alpha_{\rm GB}}\approx 0.3$ km  and 0.4 km cases have amplitudes
    that are, respectively, $\sim 10\%$  and $\sim 30\%$ smaller, compared to $\alpha_{\rm GB}=0$ (GR).
    In both panels, time is measured with
    respect to the same $t_{\rm peak}$ as in Fig.~\ref{fig:collapse_rad}.
\label{fig:charge}
}
\end{figure}

The more massive binary neutron star merger ($M=3.45\ M_{\odot}$)
is shown in Fig.~\ref{fig:collapse_rad}. After $\sim 3$--4 orbits, the binary
merges and promptly forms a black hole which rings down.
The $\ell=m=2$ component of the scalar field (bottom panel of Fig.~\ref{fig:collapse_rad})
shows similar behavior to the gravitational waves in both the inspiral and ringdown.
However, the scalar radiation is not significant enough to lead to any
noticeable dephasing in the inspiral for these parameters, and the gravitational wave 
signals for different values of $\alpha_{\rm GB}$ are indistinguishable 
on the scale of the plot, except during the ringdown.
This is consistent with the fact that the neutron stars do not have a scalar
charge, and that scalar charge only develops after the black hole forms. This
is illustrated in Fig.~\ref{fig:charge} where we show $Q_{\rm SF}$, as
measured from the average scalar field value at large distances. There it can
be seen that the scalar charge only settles to its final value $\sim 1$ ms
after the peak of the gravitational waves, while the period of gravitational waves during ringdown is $\approx
0.2$ ms.  

Perturbation
theory~\cite{Blazquez-Salcedo:2016enn,Pierini:2021jxd,Pierini:2022eim} predicts
that the real frequency of the fundamental $\ell=2$, $m=2$ quasinormal mode of
a black hole in ESGB gravity will have a smaller real frequency as the coupling
increases, and that the effect should be $<1\%$ for the values we
consider here.\footnote{
We note that the results of Refs.~\cite{Blazquez-Salcedo:2016enn,Pierini:2021jxd,Pierini:2022eim}
are obtained for Einstein-dilaton-Gauss-Bonnet gravity, 
which is equivalent to ESGB only for small values of $\phi$,
and make use of a small black hole spin expansion, and thus are only approximately
applicable to the cases studied here. 
}
Though the effect on the frequency and decay rate (imaginary frequency) of the ringdown
is small, and difficult to reliably quantify here, 
the most noticeable effect is a suppression in the overall amplitude of the 
ringdown gravitational wave signal with increasing GB coupling, as shown in the bottom panel of Fig.~\ref{fig:charge},
which occurs as the black hole develops a scalar charge.
The highest value of the ESGB coupling we consider for the prompt collapse
case is $\sqrt{\alpha_{\rm GB}}\approx 0.39$ km. This should be compared to the maximum
value for which there exists stationary black hole solutions with the same
mass and spin ($a_{\rm BH}\approx 0.8$ here), which is $\sqrt{\alpha_{\rm GB}}\approx 0.91$ km.

\begin{figure}
    \centering
    \includegraphics[width=\columnwidth,draft=false]{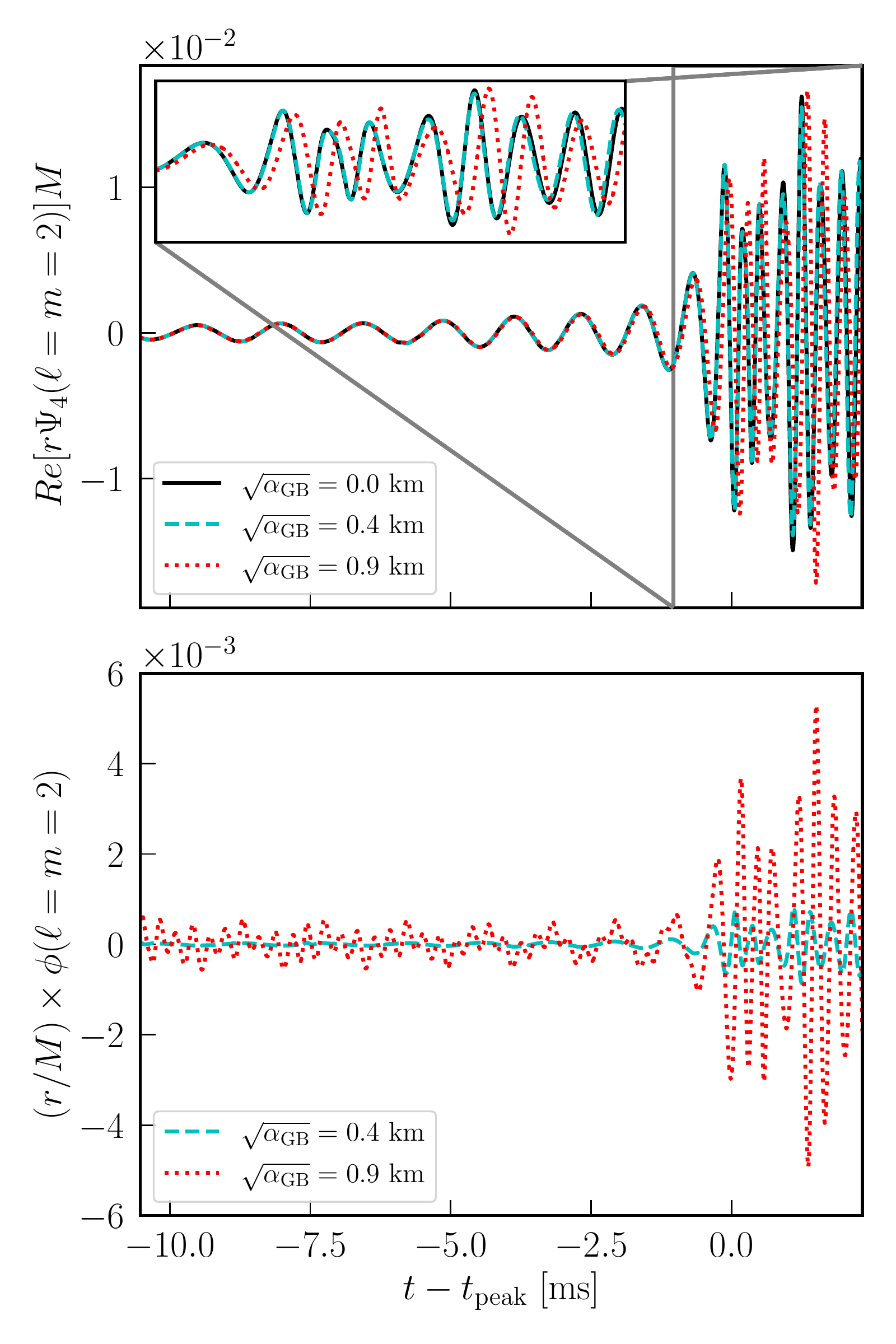}
    \includegraphics[width=\columnwidth,draft=false]{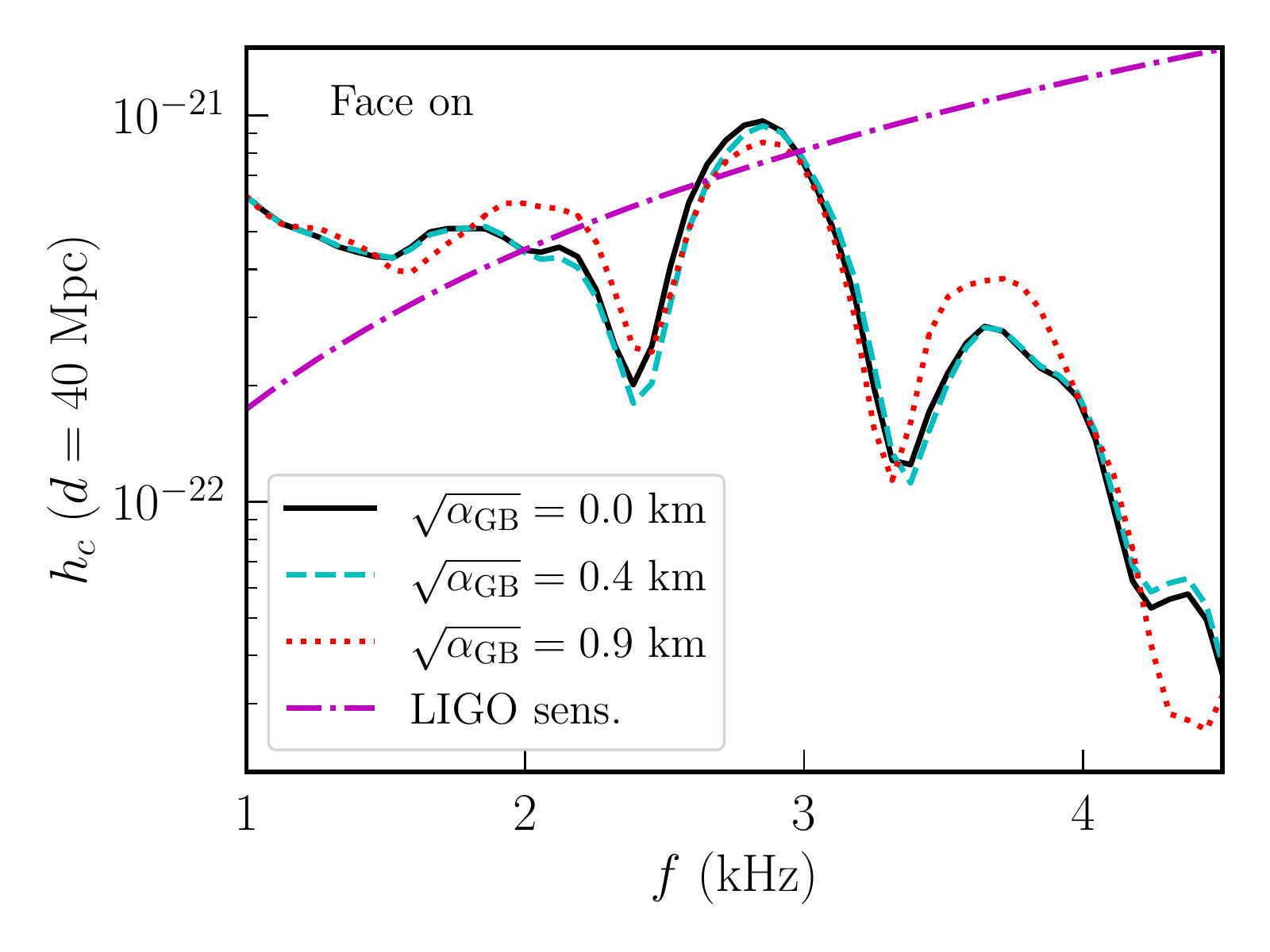}
\caption{
    As in Fig.~\ref{fig:collapse_rad}, we show the gravitational wave radiation
    (top) and scalar radiation (middle), but for a binary neutron star merger that
    forms a longer-lived remnant star (though notice the different axis scales
    compared to Fig.~\ref{fig:collapse_rad}).  For interest, we also show the
    characteristic gravitational wave strain versus frequency for these three cases (bottom), if observed
    face on at a distance of 40 Mpc, together with the advanced LIGO sensitivity
    design curve~\cite{ligo_sens}. 
\label{fig:llived_rad}
}
\end{figure}

We also consider a less massive binary neutron star merger with $M=3\ M_{\odot}$
that forms an oscillating hypermassive remnant star. We show the gravitational
and scalar radiation in Fig.~\ref{fig:llived_rad}. Without evolving to presumed
late-time black hole
formation, we are able to evolve cases with significantly larger values of 
$\alpha_{\rm GB}$ in comparison to the prompt collapse case. 
In the top panel of Fig.~\ref{fig:llived_rad}, starting slightly before merger,
and continuing to the postmerger oscillations, there is some noticeable dephasing
in the gravitational waves for the highest coupling case with $\sqrt{\alpha_{\rm GB}}\approx 0.89$ 
km.\footnote{Achieving small phase errors in the postmerger phase of binary neutron simulations is 
still an open problem, see, e.g., Ref.~\cite{Raithel:2022san}, and this comparison should be treated
as an upper bound on the gravitational wave dephasing assuming that the dominant truncation
error is similar comparing ESGB to GR simulations performed at the same resolution.
}
This difference will show up at high gravitational wave frequencies (in the kilohertz regime).
We note that a value of $\sqrt{\alpha_{\rm GB}}\approx 0.95$ km 
would exclude even a nonspinning (static) black hole solution with mass $3\ M_{\odot}$.
The scalar radiation also tracks the neutron star
oscillations evident in the gravitational waves.

In this $\sqrt{\alpha_{\rm GB}}\approx 0.89$ km case, the initial data transient from the scalar field going from zero
to nonzero in the vicinity of the star also induces measurable (yet small) oscillations
in the fundamental fluid mode of the star, known as the f-mode, which in turn cause scalar radiation during the inspiral. 
This is evident in the bottom 
panel of Fig.~\ref{fig:llived_rad}. (N.B. the higher vertical axis scale in Fig.~\ref{fig:llived_rad}
compared to Fig.~\ref{fig:collapse_rad}.) 
In this case, these f-mode oscillations are an artifact of the initial conditions, 
though similar oscillations can arise through tidal excitations, for example in 
eccentric neutron star mergers~\cite{1977ApJ...216..914T,Gold:2011df,East:2011xa,East:2012ww}.

\begin{figure}
    \centering
    \includegraphics[width=\columnwidth,draft=false]{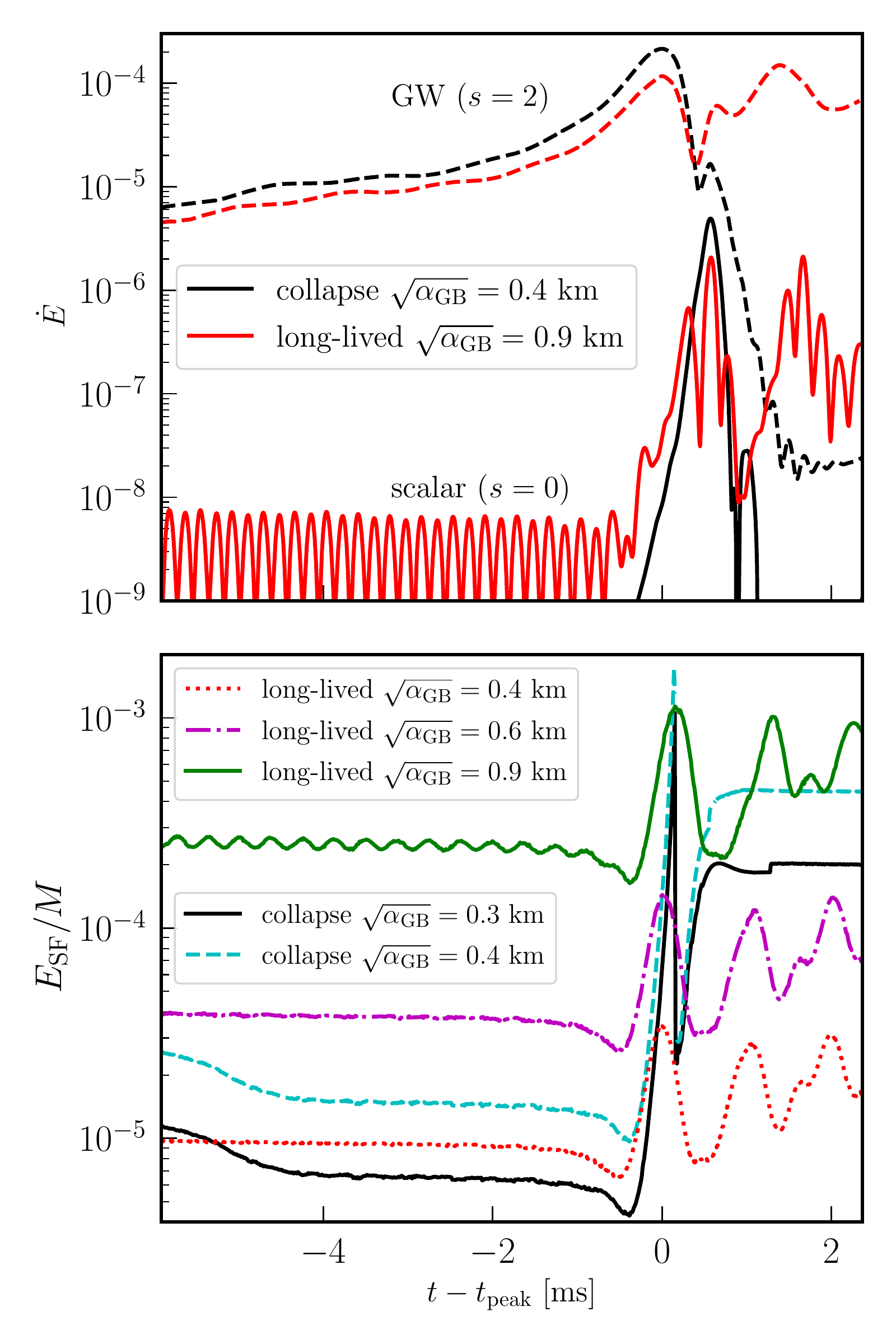}
\caption{
    Comparison of neutron star mergers with two different values of total mass:
    $M=3.4\ M_{\odot}$ (leading to prompt collapse to a black hole) and $M=3 \ M_{\odot}$ 
    (leading to a longer-lived remnant star), and different values of the GB coupling.
    \textit{Top}: The scalar (solid curves) and gravitational (dashed curves) radiation from 
         neutron star mergers that promptly collapse (black curves),
         and ones that form a longer-lived remnant (red curves). 
    \textit{Bottom}: A comparison of the canonical scalar field energy $E_{\rm SF}$ as a function of
           time for several mergers exhibiting prompt collapse or a longer-lived remnant, and
           with various values of the GB coupling.
    In all cases, the curves have been aligned in time at the gravitational wave luminosity peak.
\label{fig:power}
}
\end{figure}

We further compare the collapsing and longer-lived remnant star cases in 
Fig.~\ref{fig:power}. In both cases, the luminosity of the scalar radiation
is always subdominant to the gravitational radiation, and the former peaks after 
the latter (top panel).

\begin{figure}
    \centering
    \includegraphics[width=\columnwidth,draft=false]{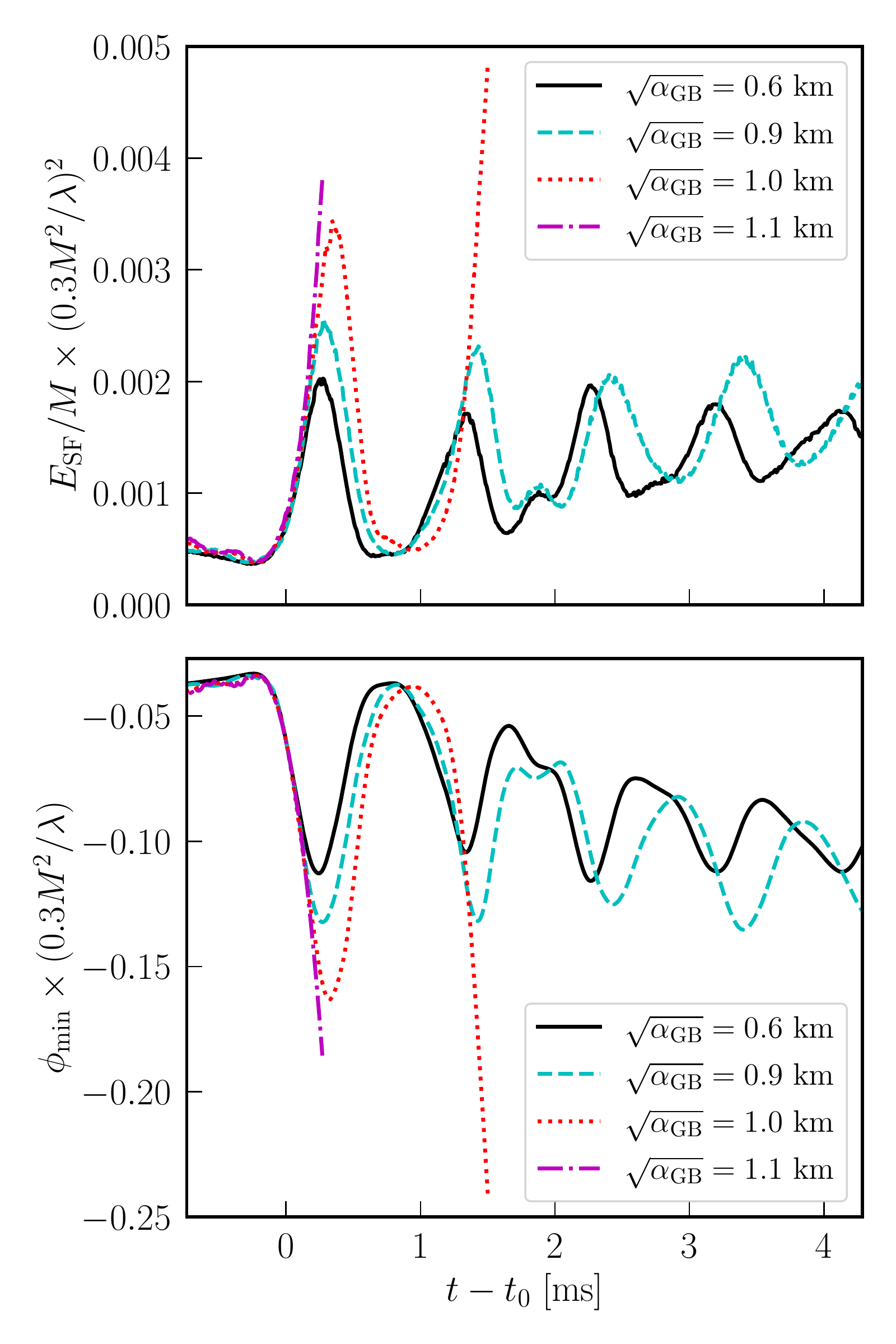}
\caption{
    The canonical scalar field energy $E_{\rm SF}$ (top panel) and the minimum
    scalar field value $\phi_{\rm min}$ at a given time (bottom panel) around
    merger for the longer-lived remnant case and different values of the GB
    coupling.  We have scaled both quantities so that they would agree with the
    highest coupling case ($\lambda/M^2=0.3$ or equivalently $\sqrt{\alpha_{\rm
    GB}}\approx 1.1$ km) assuming the test-field dependence on the coupling.
    For the cases with the two highest couplings, we were unable to continue the evolution
    past the point shown.
\label{fig:nonlinear}
}
\end{figure}

In the longer-lived remnant case, for higher values of the GB coupling than
discussed above, in particular for $\sqrt{\alpha_{\rm GB}}\gtrsim 1$ km, we
find a nonlinear enhancement in the scalar field, which reaches
values $>0.1$ (in units of the Planck mass) postmerger, and causes our evolution to breakdown before there is any
sign of collapse to a black hole. This is illustrated in
Fig.~\ref{fig:nonlinear}, where we show the scalar field energy and maximum
field magnitude for several values of the coupling. Postmerger, these
quantities oscillate with the remnant star.  After rescaling for the test-field
dependence on coupling, we can see that there is a mild nonlinear
enhancement in these quantities for $\sqrt{\alpha_{\rm GB}}\approx 0.89$ km, which
becomes strongly nonlinear for $\sqrt{\alpha_{\rm GB}}\approx 1.0$ km.  
For the highest coupling considered ($\sqrt{\alpha_{\rm GB}}\approx 1.1$ km),
the blow up in the scalar quantities happens during the first oscillation, while
for a slightly smaller value ($\sqrt{\alpha_{\rm GB}}\approx 1.0$ km) it happens
during the second oscillation.
For both of the cases, we are unable to continue the evolution further.
This could be related to a breakdown in the hyperbolicity of the ESGB equations, either
in the theory itself, or in our particular formulation and choice of gauge,
though further work would be needed to demonstrate this.
Assuming this is due to breakdown of hyperbolicity, similar to arguments
constraining $\sqrt{\alpha_{\rm GB}}$ based on the smallest observed black hole, 
the observation
of a binary neutron star postmerger without apparent anomalies
can set a constraint on $\sqrt{\alpha_{\rm GB}}\lesssim 1$ km.
However, an alternative perspective might be that ESGB is only
an approximation to a more complete gravity theory, and these cases 
may merely lie in the regime where additional corrections need to be taken into
account.

\begin{figure*}
    \centering
    \includegraphics[width=0.66\columnwidth,draft=false]{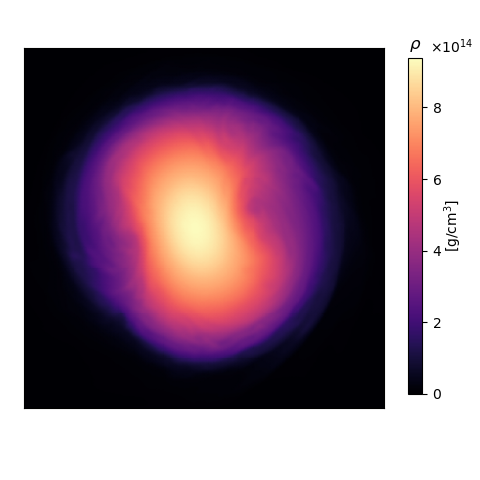}
    \includegraphics[width=0.66\columnwidth,draft=false]{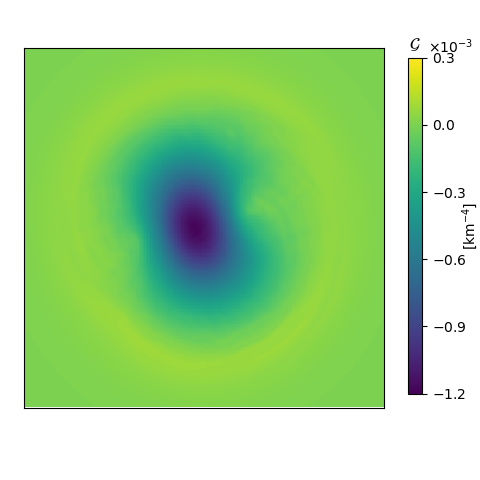}  
    \includegraphics[width=0.66\columnwidth,draft=false]{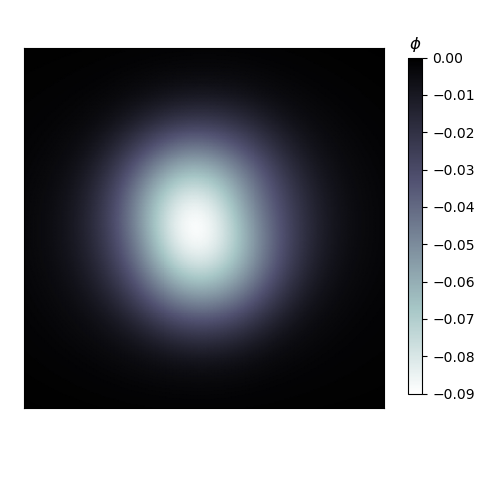}  
\caption{
    Snapshots of rest-mass density $\rho$ (left), the GB curvature scalar
    $\mathcal{G}$ (center), and the scalar field $\phi$ (right) for the case
    with $M=3 \ M_{\odot}$ and $\sqrt{\alpha_{\rm GB}} \approx 0.89$ km at a
    time following the merger (peak of the gravitational wave luminosity) 
    $t-t_{\rm peak}\approx 4$ ms.  What is shown is a zoom-in of the postmerger remnant star, where the
    coordinate distance of the linear dimension of each plot is $\approx 44$ km.
\label{fig:snapshot}
}
\end{figure*}

We show snapshots of the density, GB curvature, and scalar field around 
the time $|\phi|$ reaches a local maximum during the oscillations in the
postmerger remnant in Fig.~\ref{fig:snapshot}. 
At the center of the star, coincident with high density, the GB curvature
reaches a magnitude that is only a factor of 2 smaller than the value at the horizon of a nonspinning black hole
($\mathcal{G}\approx2\times10^{-3}$ km$^{-4}$ for a Schwarzschild
black hole with $M=3\ M_{\odot}$), though with the opposite sign. In turn, 
the scalar field is also negative with largest magnitude at the center of the star.
The maximum positive value of the GB curvature is $~\sim 4\times$ smaller in magnitude
than the maximum negative value and occurs near the surface of the star.

\subsection{Collapse of isolated hypermassive neutron stars}
One possible outcome for a binary neutron star merger is that the remnant
star undergoes a delayed collapse to a black hole, which happens only
after gravitational radiation, cooling, viscosity, and other dissipative
effects have sufficiently reduced the differential rotation and thermal
support of the star.  To cover this scenario, we consider the collapse of
a uniformly rotating hypermassive neutron star with mass 2.64 $M_{\odot}$ 
and dimensionless spin $0.7$.  The star is an unstable
equilibrium solution in GR and rapidly collapses to a black hole, with
the collapse induced either by truncation error (when $\alpha_{\rm GB}=0$) or by
the perturbation induced on the star by the modified gravity (when
$\alpha_{\rm GB}\neq 0$).

\begin{figure}
    \centering
    \includegraphics[width=\columnwidth,draft=false]{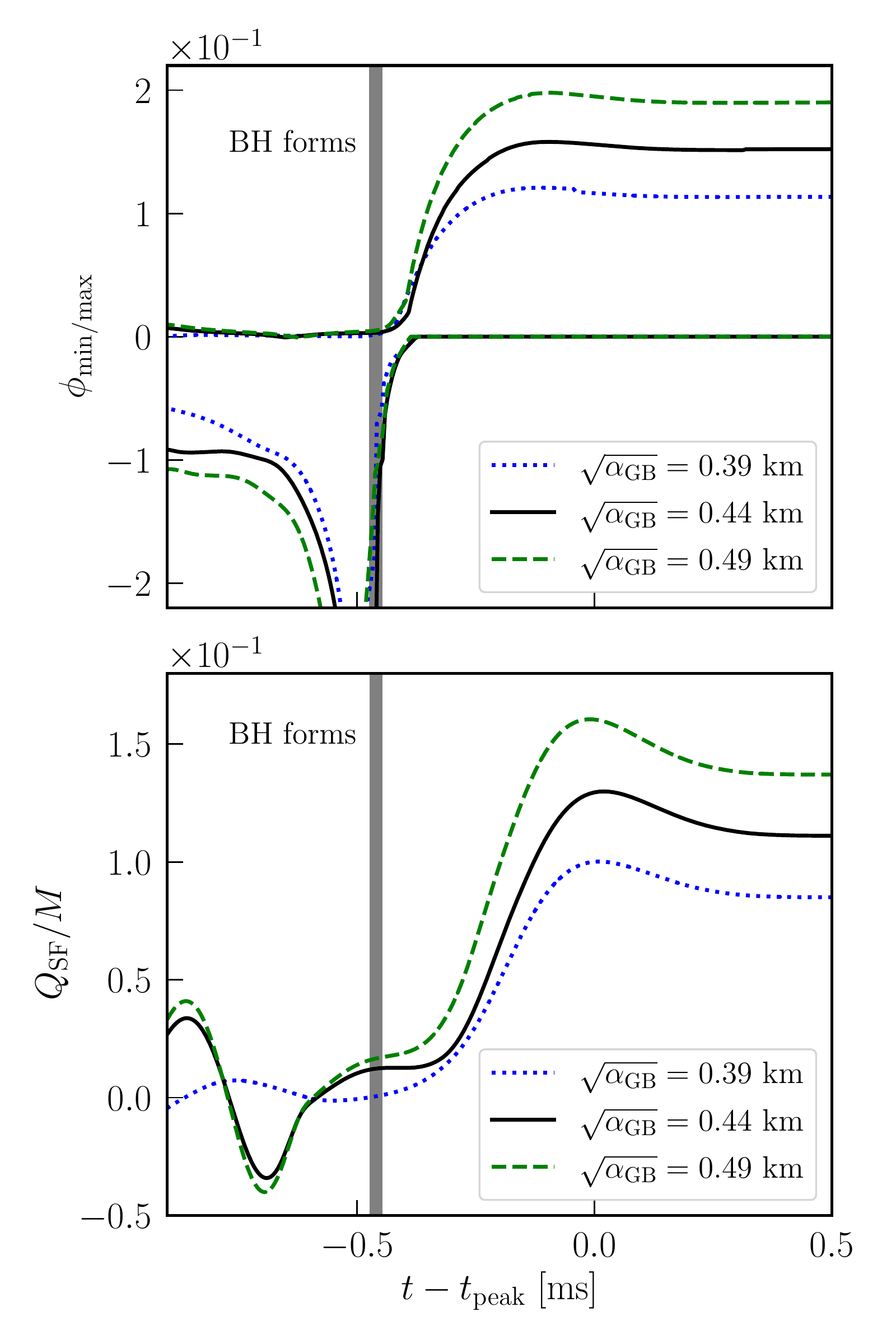}
\caption{
    The minimum and maximum value of the scalar field over the domain
     (excluding the black hole interior, top panel) and scalar charge $Q_{\rm SF}$ (bottom panel) from the collapse
     of uniformly rotating hypermassive neutron stars with different values of the GB coupling.
     The time axis has been shifted to the peak of the gravitational wave luminosity, and the gray
     band indicates the approximate time the black hole forms (measured via apparent horizon formation). 
\label{fig:rns_scalar}
}
\end{figure}

As above, in ESGB gravity the compact object develops a scalar charge as it
collapses to a black hole and rings down down to a stationary black hole
(with scalar hair) solution.  
Also as found in the neutron star mergers, the scalar field is negative, but with
growing magnitude at the center of the collapsing star, coinciding with the negative
GB curvature. However, as the black hole forms, this region is hidden,
and the magnitude of $\phi$ is peaked at a positive value in the vicinity of the black hole horizon,
which grows towards its asymptotic value as the black hole settles down.
This is illustrated in Fig.~\ref{fig:rns_scalar}.
Similar to the
prompt collapse following a neutron star merger (Fig.~\ref{fig:charge}),
the development and settling of the scalar charge to its final value takes
place over $\approx0.5$--1 ms.

\begin{figure}
    \centering
    \includegraphics[width=\columnwidth,draft=false]{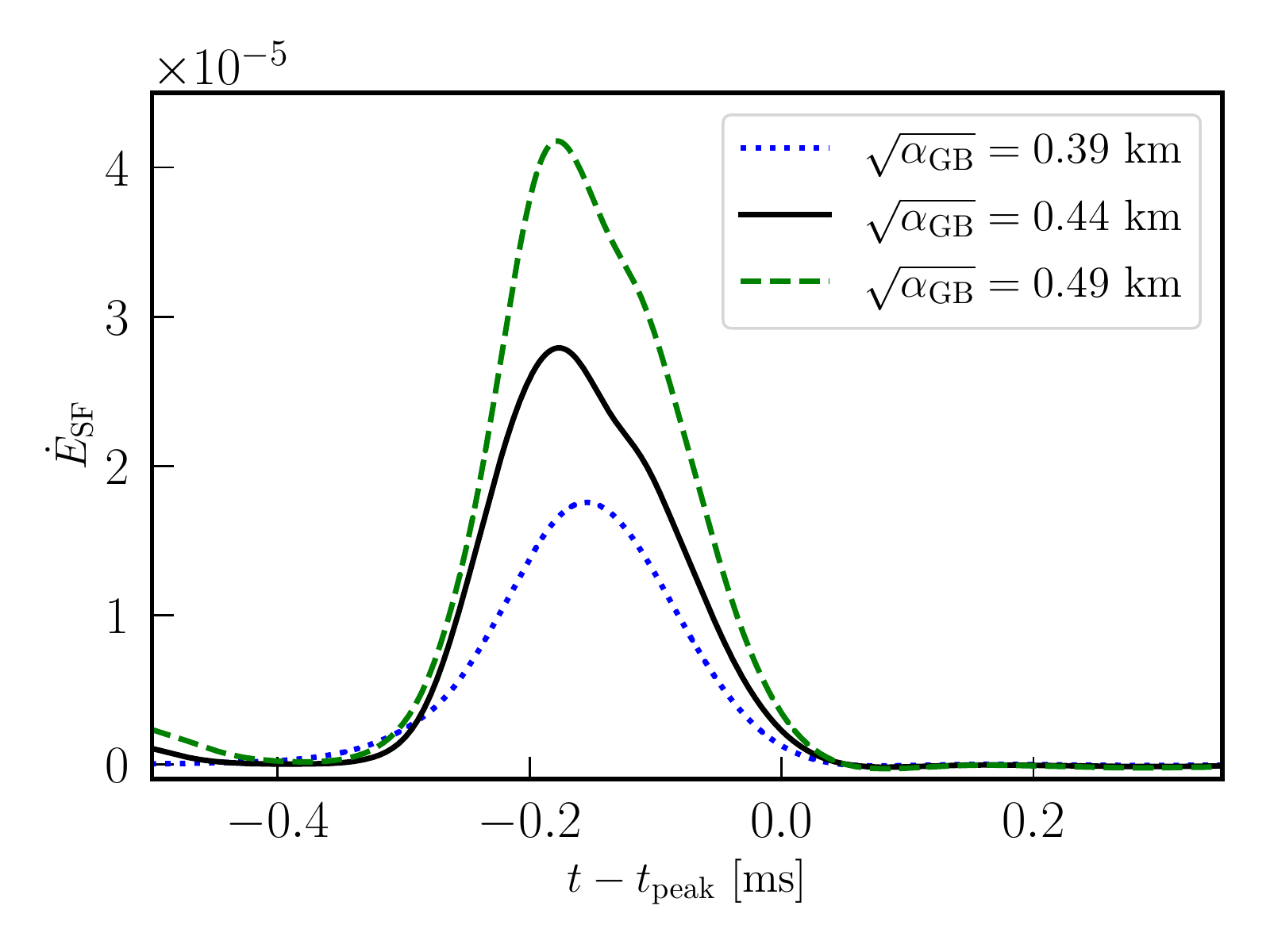}
\caption{
     The scalar luminosity from the collapse
     of uniformly rotating hypermassive neutron stars with different values of the GB coupling.
     The time axis has been shifted to the gravitational wave peak, as in Fig.~\ref{fig:rns_scalar}.
\label{fig:rns_scalar_flux}
}
\end{figure}

This transition is accompanied by a burst of scalar radiation, as shown in
Fig.~\ref{fig:rns_scalar_flux}. In this case, where the gravitational wave
radiation is almost entirely from black hole ringdown, the peak scalar
radiation slightly precedes the peak gravitational luminosity (as opposed
to the gravitational wave signal being peaked at merger, and the peak
scalar radiation following, as in Fig.~\ref{fig:power}).  

\begin{figure}
    \centering
    \includegraphics[width=\columnwidth,draft=false]{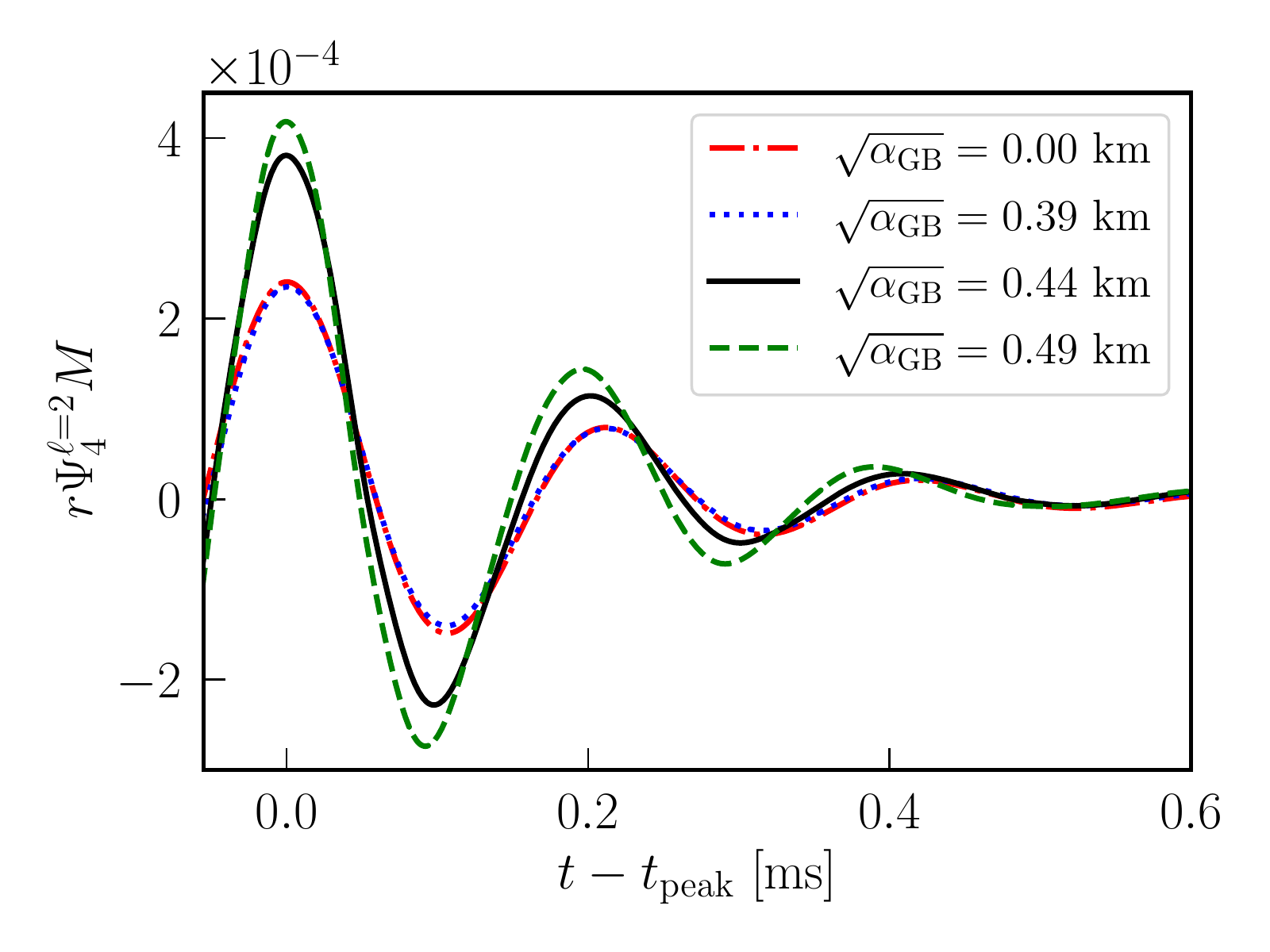}
\caption{
    Ringdown gravitational waves, in particular the $\ell=2$ ($m=0$) component of $\Psi_4$, from the collapse of a uniformly rotating hypermassive neutron star with various
    values of the GB coupling. 
     The time axis has been shifted to the gravitational wave peak, as in Fig.~\ref{fig:rns_scalar}.
\label{fig:rns_psi4}
}
\end{figure}

The gravitational wave ringdown, and its dependence on the GB coupling, is
illustrated in Fig.~\ref{fig:rns_psi4}.  There it can be seen that as the
coupling is increased, the gravitational wave amplitude also increases, which
may in part be an artifact of using as initial conditions a solution that is an
unstable stationary solution when $\alpha_{\rm GB}=0$, so the development
of a scalar field hastens the collapse to a black hole.  
We are not able to discern the expected shift in the frequency of the quasinormal mode here---in fact the trend in Fig.~\ref{fig:rns_psi4} is towards a
small decrease in period between successive peaks for larger coupling.  This is
most likely because the biggest effect of changing the GB coupling here,
as in the binary merger case above, is just
the amplitude at which different quasinormal modes (including overtones) are
excited, which could swamp a small effect on the frequency of the fundamental
mode of the final black hole. 

\section{Discussion and Conclusion}%
\label{sec:discuss}
We have used numerical evolutions of the full equations of ESGB gravity to study binary
neutron star mergers, motivated by the fact that the smaller masses of such
binaries, relative to black hole binaries, may probe modifications to GR at
smaller curvature scales.  We find that during the inspiral, there is scalar
radiation, but its amplitude is suppressed due to the fact that neutron stars
do not have scalar charge in this theory, and the impact on the gravitational wave
signal is negligible. This is true even for values of the GB coupling up to
values where there no longer exist black hole solutions with the same total
mass.  We note in passing that the scalar radiation may be enhanced if the
stars become tidally perturbed: we found that it was significantly larger for
stars that exhibited f-mode oscillations. Though here the excitation of the
oscillations was an unphysical artifact of the initial conditions, in nature this can occur
(for example)
during close encounters in neutron star binaries with orbital
eccentricity~\cite{1977ApJ...216..914T,Gold:2011df,East:2011xa,East:2012ww}.

When the neutron stars merge, the effects due to the ESGB modifications of GR
become more important. The GB curvature in the remnant star has a maximum
magnitude that is only a factor of a few less than a black hole of the same
mass, but since there is no horizon, it is peaked at the center of the star
with negative value. This gives rise to a scalar field profile that is also 
peaked at the center of the star, and with opposite sign from a black hole. 
In the case of a longer-lived remnant star, the density oscillations of the star also
cause oscillations in the scalar field and produce
scalar radiation. At larger values of the GB coupling, there is a small decrease in the frequency
of the postmerger oscillations, which in turn affects the phase of the postmerger
gravitational waves. 

In shift-symmetric ESGB, there is a minimum mass, in units of the coupling
parameter, for stationary black hole solutions, and there have been attempts
to use the putative observation of the smallest mass black holes to constrain the 
theory. It has been previously shown
that from the perspective of evolution, starting with a vacuum black hole, or collapsing to a black 
hole with mass below this
threshold leads to a breakdown in the hyperbolicity
of the evolution equations~\cite{Ripley:2019irj,Ripley:2019aqj}. Here, we find
evidence that something similar may happen in a hypermassive remnant star. In
particular, we find that for a value of the GB coupling only $\sim 30\%$ larger
than the value that would exclude a black hole of the same mass, and that is
still marginally consistent with observations, there is a strong nonlinear
enhancement in the scalar field magnitude, and a breakdown in our numerical
evolution. This is suggestive that we are near the strong-coupling regime where
the ESGB evolution equations may become elliptic, though a more detailed
analysis would be needed to establish this.

We also considered several cases where a black hole forms, both promptly
following the merger of a binary neutron star, and by considering the collapse
of a uniformly rotating hypermassive star, the latter of which approximates the
delayed collapse of a remnant after the dissipation of differential rotation.
In both cases, following the appearance of an apparent horizon, the scalar
field on the horizon and the scalar charge at large distances grows and settles
towards its final value on timescales of $\sim 0.5$--1 ms.  These cases also
allow us to self-consistently study the effect of modifications to GR on the
ringdown gravitational wave signal of newly formed black holes. Much attention
has been focused on the change in the ringdown frequency of the final black
hole in modified theories of gravity, since this is a simple quantity that can be
calculated in perturbation theory without a detailed understanding of the
merger dynamics in the modified theory. However, for the cases considered here,
the frequency shift is small, and we find that the dominant effect is actually
a change in the amplitude of the black hole perturbation that lead to the
ringdown signal. This is an additional observational signature of modified
gravity that can be potentially leveraged, but it also illustrates the
complications in ringdown tests of GR that come from including all the ways in
which the modifications will affect the ringdown signal. The gravity modification
can shift the amplitude of the ringdown, including the relative amplitude at
which different overtone modes are excited, impacting when the dominant
quasinormal mode frequency can be cleanly extracted using a finite time
interval following the peak of the gravitational wave signal, as well as
potentially changing the mass and spin of the remnant black hole compared to
GR.

Unfortunately, for binary neutron star mergers, the postmerger oscillations
and, to an even greater degree, the ringdown of the final black hole are at
kilohertz frequencies that are too high for current ground-based detectors to
be very sensitive to. So directly observing this regime will likely require
third generation detectors~\cite{Hild:2010id,LIGOScientific:2016wof} or
detectors that specifically target high frequencies~\cite{Martynov:2019gvu}.
We defer a more detailed study of the detectability of the modified gravity effects 
we find here to future work. An important aspect of assessing this would
be to determine how degenerate these effects are with different binary parameters,
and how robust they are to different choices for the unknown neutron star equation
of state.

\acknowledgments
We thank Vasileios Paschalidis and Antonios Tsokaros for their assistance
constructing the binary neutron star initial data used here.  W.E. acknowledges
support from an NSERC Discovery grant.  This research was supported in part by
Perimeter Institute for Theoretical Physics. Research at Perimeter Institute is
supported by the Government of Canada through the Department of Innovation,
Science and Economic Development Canada and by the Province of Ontario through
the Ministry of Research, Innovation and Science.  Computational resources were
provided by XSEDE under Grant TG-PHY100053, as well as by Calcul Québec
(www.calculquebec.ca) and Compute Canada (www.computecanada.ca), and the
Symmetry cluster at Perimeter Institute. F.P. acknowledges
support from NSF Grant No. PHY-2207286, the Simons Foundation, and the
Canadian Institute For Advanced Research (CIFAR).

\appendix
\section{Numerical resolution and convergence}
\label{app:conv}
\begin{figure}
    \centering
    \includegraphics[width=\columnwidth,draft=false]{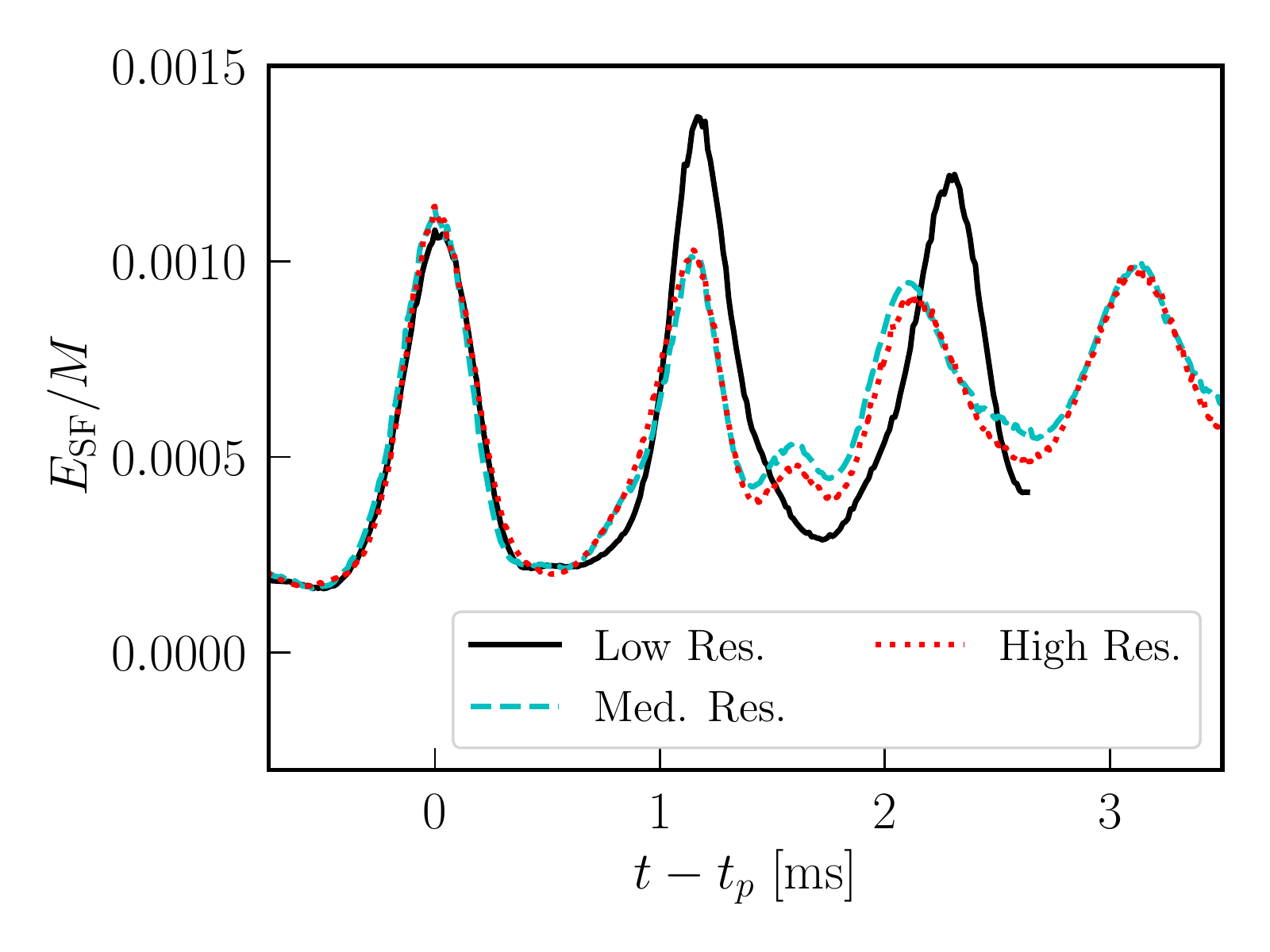}
    \includegraphics[width=\columnwidth,draft=false]{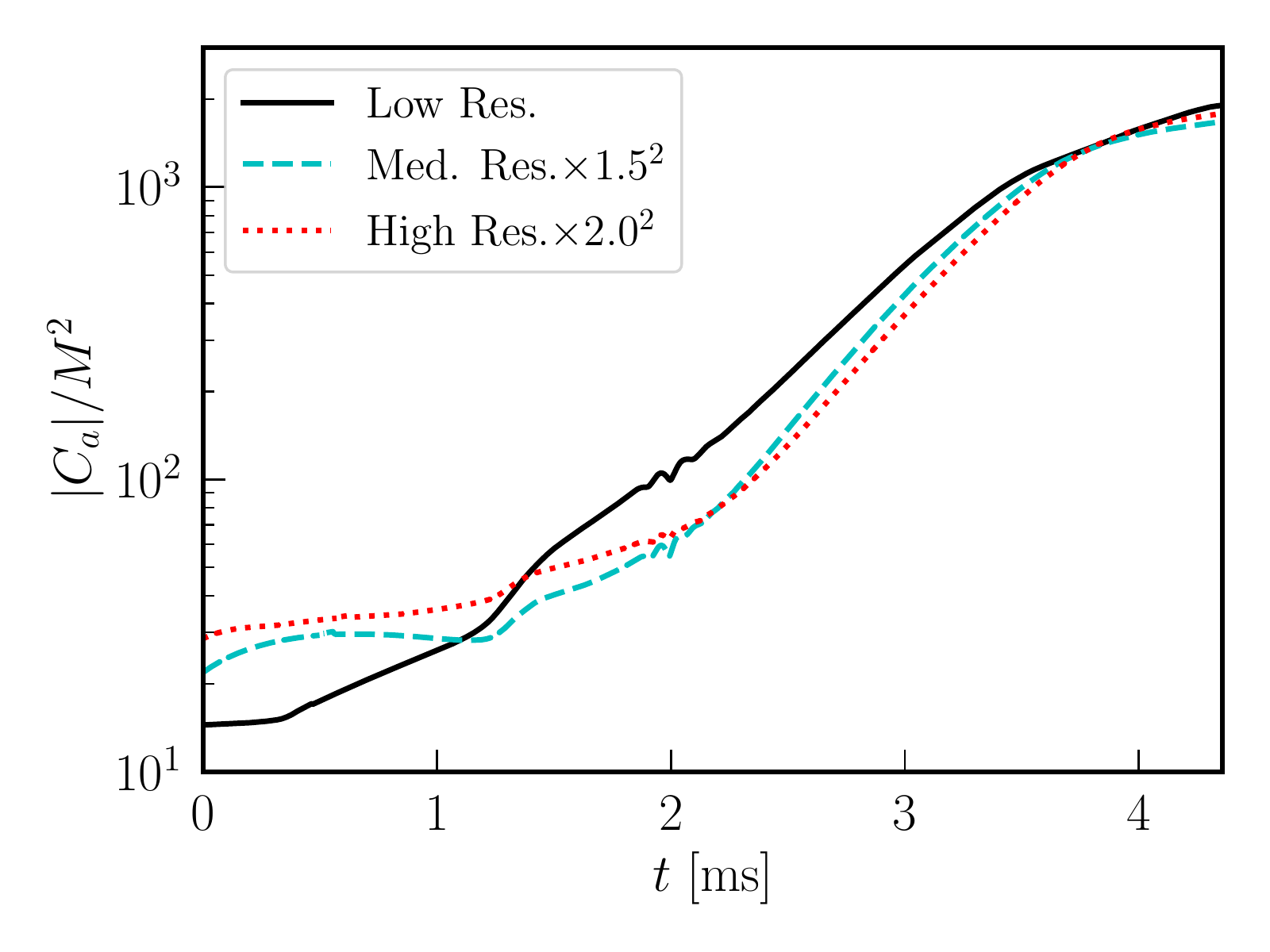}
\caption{
    Convergence results from simulations performed at three different resolutions.
    Top: The scalar field energy $E_{\rm SF}$ from a binary neutron star merger with 
    $M=3 \ M_{\odot}$ and $\sqrt{\alpha_{\rm GB}} \approx 0.89$ km. 
    Bottom: The L1 norm of the modified generalized harmonic constraint $C^a$ 
    integrated over the domain from the collapse of an isolated hypermassive neutron star
    with $\sqrt{\alpha_{\rm GB}}=0.44$ km. The values have been scaled assuming second order convergence,
    though at early times the convergence is closer to first order.
\label{fig:conv}
}
\end{figure}

For all of the binary neutron star merger cases considered in the main text, we perform
simulations with six levels of adaptive mesh refinement where the finest level
has a linear grid spacing of $dx\approx 0.05M$, and each successive level has a
grid spacing that is twice as coarse.  For the case with $M=3 \ M_{\odot}$ and
$\sqrt{\alpha_{\rm GB}} \approx 0.89$ km, we also perform a convergence study with
grid spacing that is $4/3$ and $\times 2/3$ as large. Unless otherwise stated, all
results are from the highest resolution. In the top panel of
Fig.~\ref{fig:conv}, we show how the canonical scalar field energy postmerger
(as in the top panel of Fig.~\ref{fig:nonlinear}) varies with resolution. There
it can be seen that the difference in the amplitude of the first peak in all
resolutions, and the timing and amplitude of subsequent peaks for the two
highest resolutions, is small (e.g. compared to the nonlinear effects in
Fig.~\ref{fig:nonlinear}), though there is some noticeable difference in the
lowest resolution after the oscillation.

For the simulations of the collapse of isolated hypermassive stars, we assume axisymmetry,
which makes the computational domain two-dimensional,
and use seven levels of mesh refinement with $dx\approx 0.01 M$ on the finest level.
We perform a resolution study for $\sqrt{\alpha_{\rm GB}}\approx 0.44$ km,
running simulations with grid spacing $2$ and $\times4/3$ coarser.
In the bottom panel of Fig.~\ref{fig:conv}, we show the 
norm of the modified generalized harmonic constraint~\cite{East:2020hgw}
\begin{equation}
C^a:=H^a-\tilde{g}^{bc}\nabla_b \nabla_c x^a
\end{equation}
integrated over the domain as a function of time for the three resolutions.
Though at early times the order of convergence is closer to first order,
presumably from scalar induced perturbations engaging the shock-capturing
scheme, as the star collapses to a black hole and rings down, the convergence
is consistent with approximately second order convergence (which is assumed in
the scaling of the lower panel of Fig.~\ref{fig:conv}), as expected from our
numerical scheme in the absence of shocks.

\bibliographystyle{apsrev4-1.bst}
\bibliography{mod_grav,ref}


\end{document}